\documentclass[twocolumn,showpacs,preprintnumbers,amsmath,amssymb]{revtex4}

\usepackage{graphicx}
\usepackage{dcolumn}
\usepackage{bm}

\newcommand{\microns}{\; \ensuremath{\mu\rm{m}}}

\begin{document}
\title{Time-resolved broadband analysis of slow-light propagation and superluminal transmission of electromagnetic
waves in three-dimensional photonic crystals}
\author{J. G\'omez Rivas}
\email{jaime.gomez@THz-photonics.com, www.THz-photonics.com}
\author{A. Farr\'e Benet}
\author{J. Niehusmann}
\author{P. Haring Bolivar}
\author{H. Kurz}
\affiliation{ Institut f$\ddot{u}$r Halbleitertechnik, RWTH
Aachen, Sommerfeldstr. 24, D-52056 Aachen, Germany. }

\date{\today}

\begin{abstract}
A time-resolved analysis of the amplitude and phase of THz pulses
propagating through three-dimensional photonic crystals is
presented. Single-cycle pulses of THz radiation allow measurements
over a wide frequency range, spanning more than an octave below, at
and above the bandgap of strongly dispersive photonic crystals.
Transmission data provide evidence for slow group velocities at the
photonic band edges and for superluminal transmission at frequencies
in the gap. Our experimental results are in good agreement with
finite-difference-time-domain simulations.
\end{abstract}
\maketitle

\section{Introduction}

Electromagnetic waves propagate in straight trajectories in
homogeneous materials. This simple propagation changes when the wave
encounters inhomogeneities and is scattered. Interference of the
scattered waves can modify significantly their propagation. The
careful engineering of the inhomogeneities may lead to the
inhibition of the propagation in certain directions and to the
enhancement in others. It is also possible to
speed-up~\cite{Steinber:93,Spielmann:94,Longhi:01,Hache:02,Munday:02,Solli:03,Galli:04}
or
slow-down~\cite{Tarhan:95,Imhof:99,Vlasov:99,Bayindir:00,Notomi:01,Galli:04}
the propagation of the wave and even to localize
it~\cite{Wiersma:97}. The full control of the propagation of
electromagnetic waves is intensively pursued since this control will
lead to new concepts in numerous fields such as information
processing~\cite{Noda:03}, laser physics~\cite{Painter:99},
bio-sensing~\cite{Alivisatos:04} and quantum
optics~\cite{Lodahl:04}.

One of the most prominent candidates for the control of the
propagation of electromagnetic waves are photonic crystals. Photonic
crystals are periodic structures of two or more materials with
different refractive indices~\cite{Yablonovitch:87,John:87}.
Interference of waves scattered at different lattice planes of the
crystal determine its optical properties. Depending on the structure
and if the scattering is strong enough a photonic band gap (PBG)
might be created. A PBG is a frequency range in which no optical
modes are allowed and, consequently, the propagation of waves in
this range is forbidden. Particulary attractive is the extraordinary
optical dispersion that photonic crystals exhibit, which leads to a
pronounced variation of the wave's group velocity. The resonant
scattering at the band edges of the gap gives rises to a strong
reduction of the group velocity and a significant group velocity
dispersion~\cite{Imhof:99}. This slow propagation, which increases
the interaction time between the wave and the crystal, constitutes
the basis for the development of sensors for gases and
bio-molecules~\cite{Alivisatos:04} and more sensitive non-linear
components~\cite{Soljacic:04}. A wave with a frequency in the PBG
incident on a photonic crystal is reflected. However, if the crystal
has a finite thickness, a fraction of the wave's intensity is
transmitted. While the transmitted intensity decreases exponentially
with the crystal thickness, the transmission time is independent of
this thickness, which may lead to superluminal transmission.
Tunneling processes through photonic structures are of great
scientific and technological
interest~\cite{Chiao:97,Nimtz:03,Stenner:03,Nimtz:04}. Among the
reasons driving this interest are the similarities with the
tunneling of quantum particles and the ongoing intense discussions
on the propagation in the superluminal
regime~\cite{Stenner:03,Nimtz:04}.

In spite of the fundamental and technological relevance of dynamic
properties of electromagnetic waves in photonic crystals most of the
investigations so far focus only on stationary measurements. This
focus is mainly imposed by the complexity of phase sensitive
techniques at high frequencies. Reports on the group velocity of
waves in optical photonic crystals are thus seldom and limited to
narrow frequency ranges~\cite{Munday:02}, to weakly scattering
photonic crystals~\cite{Tarhan:95,Imhof:99,Vlasov:99,Longhi:01}, or
to low dimensional photonic structures
~\cite{Steinber:93,Spielmann:94,Longhi:01,Hache:02,Solli:03,Galli:04}.

In this article we present a broadband time domain investigation of
the propagation of electromagnetic waves in
strongly-scattering-three-dimensional photonic crystals of variable
size. In our investigation we use terahertz time-domain
spectroscopy, a technique capable to generate ultrashort broadband
pulses of THz radiation and to detect the amplitude and the phase of
these pulses. Through the direct access of amplitude and phase an
unprecedented precision in the analysis of the propagation of
electromagnetic waves in photonic crystals is achieved. The
extinction coefficient is obtained from the exponential decrease of
the transmitted amplitude with the thickness of the crystal. From
the phase analysis of the transmitted wave the wave number, the
effective refractive index and the group velocity are derived. Very
strong dispersion at the edges of the gap and anomalous dispersion
at frequencies in the gap are observed. Even in a photonic crystal
with a thickness of only four unit cells the group velocity is
dramatically reduced at the band edges by almost a factor 15 with
respect to the speed of light in vacuum, while anomalous dispersion
in the gap gives rise to superluminal transmission.

\section{Samples}

\begin{figure}[b]
\centerline{\scalebox{.4}{\includegraphics{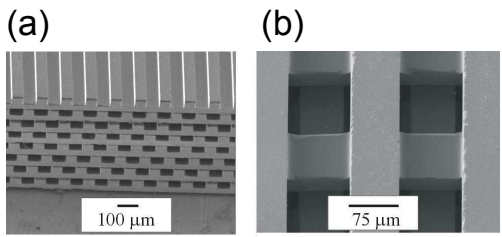}}} \caption{
Scanning electron microscope photographs of a Si layer-by-layer
photonic crystal with a thickness of 4 unit cells. Photograph (a) is
a side view, while (b) is an upper view.} \label{SEM}
\end{figure}

The photonic crystals investigated are formed by intrinsic silicon
and air. Intrinsic silicon has a very large refractive index
$n\simeq 3.4$, and virtually no absorption at terahertz frequencies.
Photonic crystals are fabricated by micromachining thin wafers of
silicon, and piling them to form a layer-by-layer
structure~\cite{Chelnokov:97,Gonzalo:02}. Such a structure consists
of a transverse stacking of arrays of dielectric rods with a
face-centered-tetragonal lattice symmetry~\cite{Ho:94,Sozuer:94}.
This type of crystal structure exhibits a very large photonic band
gap~\cite{Ho:94}. The micromachining of our crystal was done with a
programmable dicing saw, as described in more detail in
Ref.~\cite{Gonzalo:02}: parallel grooves spaced by $185 \microns$
were diced on one side of a wafer with a thickness of $138 \pm 3
\microns$. The thickness of the grooves is $110 \microns$ and the
depth one half of the thickness of the wafer, i.e., $69 \microns$.
Perpendicular grooves to these with the same thickness and depth
were diced in the opposite side. This method produces a square grid
of apertures. To form the photonic crystal structure several of
these grids are piled. Consecutive grids are shifted by one half of
the period and two of them form a single unit cell in the stacking
direction. The dimension of this unit cell is thus $276 \microns$.
The calculation of the band structure of this specific photonic
crystal predicts a PBG centered at 0.5 THz. Figures~\ref{SEM}(a) and
(b) are scanning electron microscope images of a layer-by-layer
structure with a thickness of 4 unit cells. Figure~\ref{SEM}(a) is a
side view of the structure along the stacking direction, while (b)
is a top view of the same photonic crystal.

\section{Terahertz transmission through photonic crystals}

\subsection{experimental setup}

\begin{figure}
\centerline{\scalebox{1.2}{\includegraphics{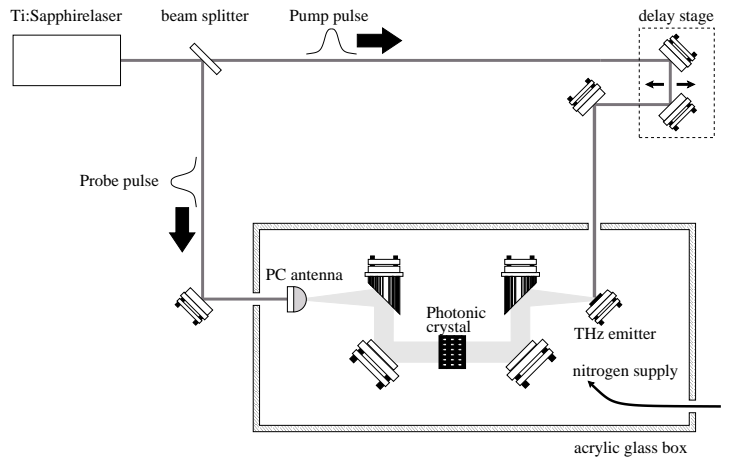}}}
\caption{Schematic representation of the experimental setup.
Terahertz pulses are generated on an InGaAs surface with a laser
pulse. These pulses are collimated and incident normal to the
crystal's surface. The transmitted amplitude is focused onto a
photoconductive antenna which is gated by a second laser pulse. By
varying the path length difference between the two optical paths the
THz amplitude is detected as a function of time. To avoid water
vapor absorption the THz setup is enclosed in a box which is
continuously fluxed with nitrogen.} \label{setup}
\end{figure}

For the experiments we use a THz time-domain
spectrometer~\cite{Exter:90}. This setup is schematically depicted
in~Fig.~\ref{setup}. A train of femtosecond pulses from a
Ti:sapphire laser is split to generate and detect single-cycle
terahertz pulses. One beam, the so-called pump beam, is incident on
an InGaAs surface field emitter, generating THz radiation. The THz
pulses are collected and collimated by a parabolic mirror placed at
the focal distance to the emitter. This THz beam is incident normal
to the surface of the crystal, and the transmitted amplitude is
focused with a parabolic mirror onto a photoconductive antenna. The
antenna is gated with the pulses from the second beam of the
Ti:sapphire laser. By means of a computer controlled delay stage,
the length difference between the generation and detection optical
paths is varied and the THz field amplitude is detected as a
function of time with sub-picosecond time resolution. Water vapor
absorbs THz radiation, therefore, the THz setup is enclosed in a box
which is continuously fluxed with nitrogen gas.

Even though the measurements were done at only one direction, it has
been demonstrated by Imhof {\it et al.} that 1D models are not able
to explain the measurements of the group velocity near the band
edges of 3D photonic crystals~\cite{Imhof:99}.

\subsection{Transmission measurements}
\label{measurements}

\begin{figure}
\centerline{\scalebox{.4}{\includegraphics{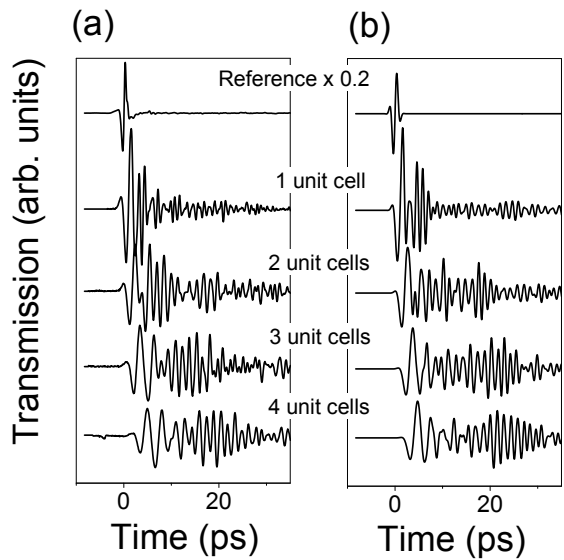}}}
\caption{ Terahertz pulses transmitted through photonic crystals
with different thickness. The thickness is indicated in the figure
as the number of unit cells (1 unit cell = $276 \mu{\rm m}$).
Measurements are plotted in (a), while the corresponding FDTD
simulations are displayed in (b). For clarity a vertical offset has
been introduced to the different pulses. The uppermost pulses are
the reference, i.e., the instrumental response of our setup.}
\label{pulses}
\end{figure}

We have measured the transmission of THz pulses through photonic
crystals with different thickness ranging from 1 to 4 unit cells.
The experimental transients of the transmitted field amplitude are
plotted in Fig.~\ref{pulses}(a). The reference pulse, i.e., the
setup response, is also plotted in the same figure.

The measurements are compared with numerical simulations using the
finite-difference-time-domain FDTD method. These simulations are
plotted in Fig.~\ref{pulses}(b), and were performed with a
commercial software (FullWave by RSoft). The spatial resolution of
the simulations is 3 $\mu{\rm m}$ and the temporal resolution is 5
fs. The incident field was approximated to a plane wave by using a
Gaussian field distribution with a width much larger than the size
of the simulation region. The temporal shape of the input THz field
was a single-cycle pulse, with a center wavelength of 500 $\mu{\rm
m}$. The transmitted field was calculated at a distance of 2.3 mm
from the source.

As can be seen in Fig.~\ref{pulses}, the transmitted pulse
experiences a significant dispersion growing with the thickness of
the crystal. Signatures of the PBG formation become directly visible
in the time domain data. Particulary in the thickest crystals, low
frequencies, belonging to the lower photonic crystal band, and high
frequencies, associated to higher bands, are separated in time
domain.

To analyze the transmission we calculate the complex Fourier
transform of the time-domain transmission through the crystal and
normalized it by the reference measured without the sample. This
transformation leads to a transmission $T(\nu)$ of the form
\begin{equation}
T(\nu)=t(\nu)e^{i\Delta\phi(\nu)} \; . \label{T}
\end{equation}
The modulus $t(\nu)$ is the spectrally resolved field transmission
coefficient, and the argument $\Delta \phi(\nu)$ is the phase
difference between the transmission through the crystal and the
reference. Equation (\ref{T}) can be written as
\begin{equation}
T(\nu)=t(\nu)e^{i k_{\rm eff}(\nu)L} \;, \label{T2}
\end{equation}
where $k_{\rm eff}$ is an effective propagation constant for the
transmission through the crystal and $L$ is the thickness of the
crystal. The effective propagation constant is given by
\begin{equation}
k_{\rm eff}(\nu)= k_{\rm eff}'(\nu) + i k_{\rm eff}''(\nu)=\frac{2
\pi}{\lambda}[n_{\rm eff}(\nu)-1+ i\kappa_{\rm eff}(\nu)]\;,
\label{kc}
\end{equation}
where $n_{\rm eff}$ is the effective refractive index and
$\kappa_{\rm eff}$ the extinction coefficient of the crystal. The
transmission is thus proportional to
\begin{equation}
T(\nu)\propto e^{-\frac{2\pi}{\lambda}\kappa_{\rm eff}(\nu)L}
e^{i\frac{2 \pi}{\lambda}(n_{\rm eff}(\nu)-1)L} \;. \label{Trans}
\end{equation}
The amplitude transmitted through the crystal decreases
exponentially with the crystal thickness
\begin{equation}
t(\nu)\propto e^{-\left[\frac{2\pi}{\lambda}\kappa_{\rm
eff}(\nu)L\right]}\;, \label{transamp}
\end{equation}
while the phase difference is given by
\begin{equation}
\Delta \phi(\nu) = \frac{2 \pi}{\lambda} (n_{\rm eff}(\nu)-1)L\;.
\label{deltaphi}
\end{equation}
It is important to mention that the transmission is measured outside
the photonic crystal. Therefore the propagation constant has a real
component or wave number $k_{\rm eff}'(\nu)$ even at frequencies
within the gap. In section~\ref{phaseanalysis} the analysis of the
phase and the dispersion of the photonic crystals are discussed in
detail.

\begin{figure}[b]
\centerline{\scalebox{.35}{\includegraphics{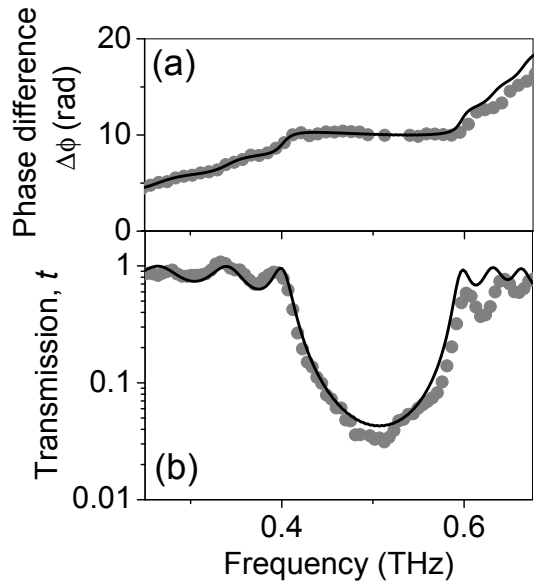}}}
\caption{(a) phase difference between the transmission through a
photonic crystal with a thickness of 3 unit cells and a reference
measurement. (b) Amplitude transmission spectrum normalized by the
reference. The wave incidences normally to the crystal surface. The
solid lines correspond to a FDTD simulation.}\label{trans}
\end{figure}

The amplitude transmission coefficient $t$ and the phase difference
$\Delta \phi$ of a photonic crystal with a thickness of 3 unit cells
are plotted in Fig.~\ref{trans}. The circles in this figure
correspond to the measurements and the solid lines are results of
the FDTD simulation. All the crystal parameters in the simulation,
i.e., structure, dimensions and refractive index, are known, and the
remarkably good agreement between simulation and measurement is
obtained without adjusting any parameter. The transmission
(Fig.~\ref{trans}(b)) drops strongly in the frequency range defined
by the gap, i.e., 0.4-0.6 THz. At these frequencies the phase
difference (Fig.~\ref{trans}(a)) is nearly constant. The
interference pattern that is apparent by the oscillations in the
transmission at low and high frequencies are Fabry-Perot resonances
due to multiple reflections at the crystal interfaces.

The extinction coefficient $\kappa_{\rm eff}$ describes the
attenuation of the wave amplitude as it propagates through the
medium. In photonic crystals made of non-absorbing constituents,
such as those investigated here, the extinction of the transmitted
amplitude at frequencies within the gap is due to destructive
interference of multiply scattered waves. At these frequencies the
wave is Bragg reflected and only a small fraction is transmitted
through the crystal. According to Eq.(\ref{transamp}) the amplitude
decreases exponentially with the crystal thickness and with an
attenuation length $L_{\rm a}=\frac{\lambda}{2\pi\kappa_{\rm eff}}$.
This attenuation length is a direct measure of the photonic strength
of the crystal. Weakly scattering photonic crystals have attenuation
lengths of the order of tens of unit cells, while $L_{a}$ in
strongly scattering crystals is on the order of a few unit cells.

\begin{figure}
\centerline{\scalebox{.4}{\includegraphics{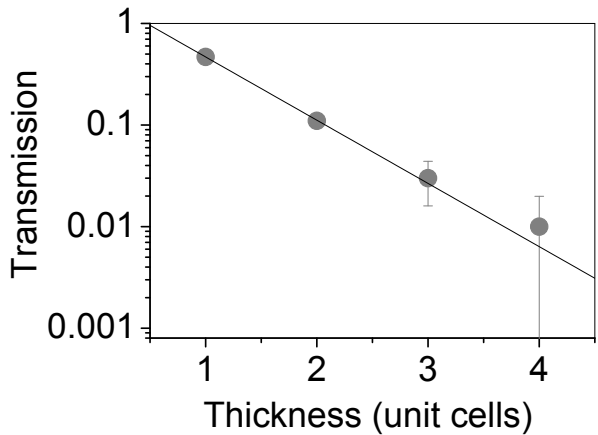}}}
\caption{Amplitude transmission at the center frequency of the gap
as a function of the thickness of the crystal. The solid line is an
exponential fit to the measurements.}\label{transvsl}
\end{figure}

Figure~\ref{transvsl} depicts the minimum transmission centered at
the gap frequency of 0.5 THz as a function of the thickness of the
crystal. The line in Fig.~\ref{transvsl} is an exponential fit to
the measurements from which we obtain an amplitude attenuation
length $L_{\rm a}$ of $195 \pm 20 \microns$.  This attenuation
length is even smaller than one unit cell ($\simeq 0.7$ times the
lattice constant), indicating an extraordinarily strong scattering
in Si layer-by-layer structures~\cite{Ozbay:96,Lin:98}.

\begin{figure}
\centerline{\scalebox{.4}{\includegraphics{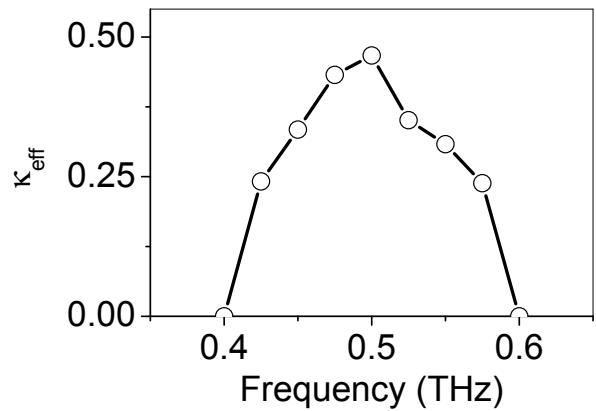}}}
\caption{effective extinction coefficient of Si photonic
crystals.}\label{keff}
\end{figure}

The extinction coefficient is represented in Fig.~\ref{keff} for
different frequencies. Radiation with a frequency outside the gap
can propagate through the crystal being only weakly reflected at the
interfaces. For these frequencies we have $\kappa_{\rm eff} = 0$. At
frequencies within the gap the transmitted amplitude decreases
exponentially with $L$ leading to a positive $\kappa_{\rm eff}$. The
largest value of $\kappa_{\rm eff}$, or equivalently the smallest
$L_{a}$, is at the central frequency of the gap.

\subsection{Phase analysis}
\label{phaseanalysis}

\begin{figure}
\centerline{\scalebox{.4}{\includegraphics{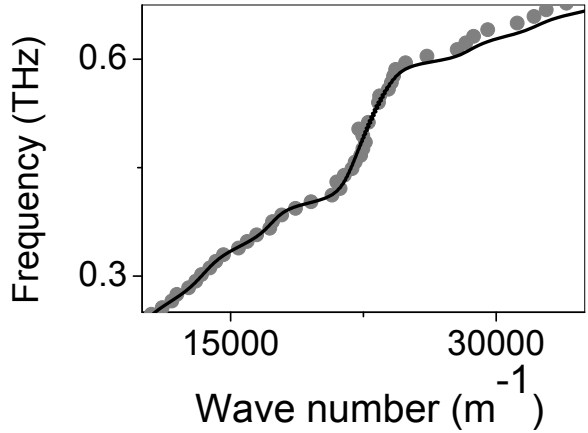}}}
\caption{ Frequency versus wave number along the stacking direction
of a Si photonic crystal with a thickness of 3 unit cells. The solid
line is a FDTD simulation.}\label{disrel}
\end{figure}

The dispersion properties of the electromagnetic wave propagating
through the crystal were obtained as described in
Refs.~\cite{Robertson:92,Ozbay:94} and in the previous section. The
measured phase difference between the transmission through the
photonic crystal and the reference was used to calculate the wave
number
\begin{equation}
k'(\nu) =\frac{2\pi}{\lambda}n_{\rm eff}(\nu)=\frac{\triangle
\phi}{L}+\frac{\omega}{{\rm c}_{0}} \;, \label{k}
\end{equation}
where $L$ is the thickness of the photonic crystal, ${\rm c}_{0}$ is
the speed of light in vacuum and $\omega = 2\pi\nu$ the angular
frequency. The measured wave number of a layer-by-layer structure of
3 unit cells is shown in Fig.~\ref{disrel} with circles. The solid
line in this figure represents the data from the FDTD simulations.
The difference between measurement and simulation at high
frequencies can be ascribed to small imperfections and misalignments
in the structure.

\begin{figure}[b]
\centerline{\scalebox{.4}{\includegraphics{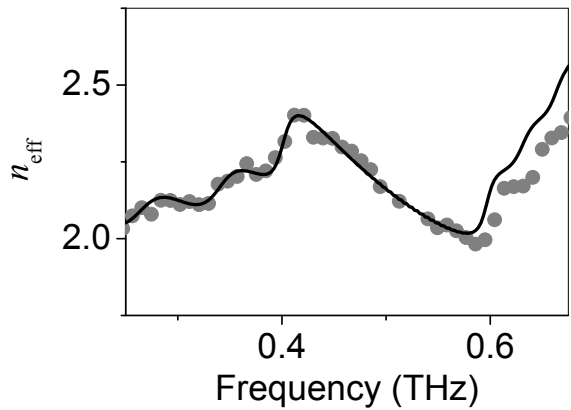}}}
\caption{Frequency dependence of the effective refractive index of a
Si photonic crystal with a thickness of 3 unit cells. The solid line
is a FDTD simulation.}\label{neff}
\end{figure}

The effective refractive index of the crystal $n_{\rm eff}$, defined
as $n_{\rm eff}(\nu)=\frac{\lambda}{2\pi}k'(\nu)$, is plotted in
Fig.~\ref{neff} versus the frequency. The FDTD simulation is
represented in the same figure with a solid line. There is a
distinct increase of $n_{\rm eff}$ at the low frequency band edge,
i.e., 0.4 THz, followed by a remarkable decrease in the gap. At the
high frequency band edge, i.e., 0.6 THz, $n_{\rm eff}$ rises again.
As we are going to see, the increase of the effective refractive
index at the band edges gives rise to a strong reduction of the
group velocity, while the anomalous dispersion in the gap leads to
superluminal transmission.

To calculate the group velocity $v_{\rm g}$, we use the standard
definition $v_{\rm g}= \frac{L}{\tau_{g}}$, where
$\tau_{g}=\frac{\partial\Delta\phi}{\partial\omega} + \frac{L}{{\rm
c}_{0}}$ is the group time delay. With the expressions of $k'$ and
$n_{\rm eff}$, the effective group velocity can be written in the
familiar form~\cite{Jackson}
\begin{equation}
v_{\rm g}(\nu)=\frac{{\rm c}_{0}}{n_{\rm eff}+\omega \partial n_{\rm
eff}/\partial \omega}\;. \label{vg2}
\end{equation}
The dispersion in $v_{\rm g}$ introduced by the photonic crystal
is described by the second term of the denominator, i.e., the
derivative of $n_{\rm eff}$ with respect to the frequency.

\begin{figure}
\centerline{\scalebox{.4}{\includegraphics{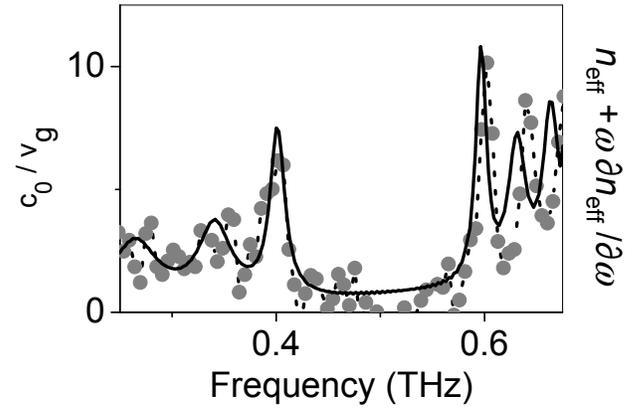}}} \caption{
Inverse of the group velocity multiplied by the speed of light in
vacuum versus the frequency of a Si photonic crystal with a
thickness of 3 unit cells. The dashed line is a guide to the eye.
The solid line is a FDTD simulation.}\label{vg}
\end{figure}

The inverse of the group velocity multiplied by ${\rm c}_{0}$, i.e.,
the denominator of Eq.~\ref{vg2}, is plotted in Fig.~\ref{vg} with
circles, and the corresponding FDTD simulation with a solid line.
The large dispersion at the band edges due to the resonant
scattering gives rise to a strong reduction of the group velocity.
The oscillations of the group velocity at low and high frequencies
that are clearly visible in the simulation are due to the multiple
reflections at the crystal interfaces. The anomalous dispersion at
frequencies in the gap, which is characterized by the negative value
of $\partial n_{\rm eff}/\partial \omega$, increases the group
velocity as defined by Eq.~\ref{vg2}. If the crystal is thick enough
this group velocity may be larger than the speed of light in vacuum.
This result can be appreciated in Fig.~\ref{vg}, where the average
value of ${\rm c}_{0}/v_{\rm g}$ in the gap of a crystal with a
thickness of only 3 unit cells is $0.4 \pm 0.8$.

\begin{figure}
\centerline{\scalebox{.4}{\includegraphics{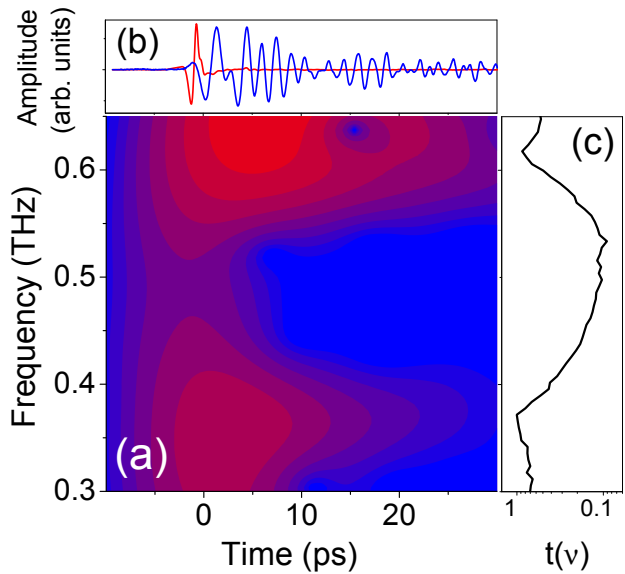}}}
\caption{(a) Short time Fourier transform of the THz transmission
through a Si photonic crystal with a thickness of 2 unit cells. The
transmitted pulse and the reference are represented in figure (b)
with a blue and red lines respectively. The amplitude transmitted
through the crystal normalized by the reference is plotted in figure
(c) as a function of the frequency.}\label{STFT}
\end{figure}

Although the analysis presented above is the most accurate for the
study of the dispersion characteristics of photonic structures,
time-frequency representation methods are more intuitive and might
be useful in this study~\cite{Capus:03}. As an example of these
methods, we plot in Fig.~\ref{STFT} the short time Fourier transform
STFT~\cite{Gamble:99} of the transmission through the photonic
crystal with a thickness of 2 unit cells. The STFT is calculated by
applying a window or a band pass filter to the time domain signal
and Fourier transforming the signal within this window to obtain the
spectral contain for a certain time delay. The process is repeated
by shifting the time window over the entire time interval of the
measurement. In this way temporal and spectral information are
simultaneously retrieved. The contour plot of Fig.~\ref{STFT}(a) is
the transmitted amplitude through the photonic crystal, where the
blue stands for no transmission and red for high transmission, as a
function of time and frequency, calculated with a Hanning time
window 8 ps wide. The THz transients are plotted in
Fig.~\ref{STFT}(b), where the red curve is the transmission through
the crystal and the blue line corresponds to the reference. The
amplitude transmission coefficient is plotted in Fig.~\ref{STFT}(c).
Consistently with the phase analysis, the time delay is the largest
at the band edges, where the group velocity reaches its minimum
value. At frequencies within the gap the transmission decreases
significantly and the time delay becomes smaller.

\begin{figure}
\centerline{\scalebox{.4}{\includegraphics{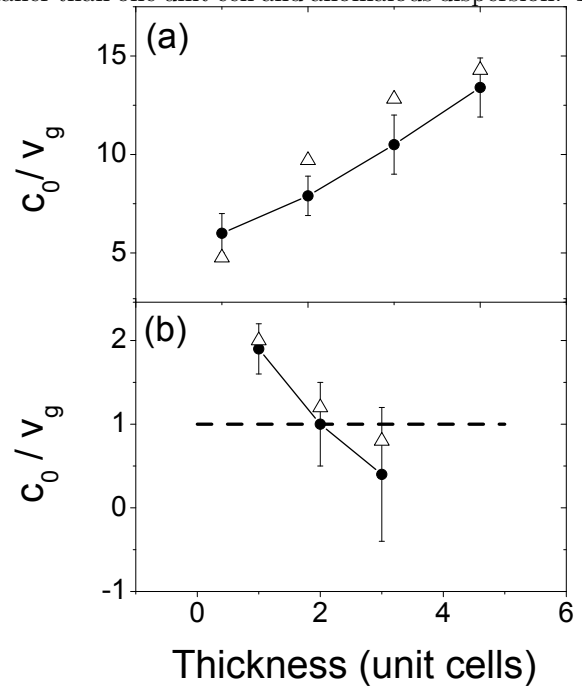}}}
\caption{Inverse of the group velocity multiplied by the speed of
light in vacuum as a function of the thickness of the photonic
crystal in unit cells. The circles are measurements, while the open
triangles correspond to FDTD simulations. (a) corresponds to the
high frequency edge of the gap, while the data in (b) are at
frequencies in the gap. The dashed line in this figure indicates the
speed of light in vacuum. The solid lines are guides to the
eye.}\label{vgvsl}
\end{figure}

Figure~\ref{vgvsl} summarizes the salient features of the group
velocity and presents its dependence on the thickness of the
crystal. Fig.~\ref{vgvsl} (a) displays the inverse of $v_{\rm g}$
multiplied by ${\rm c}_{0}$ at the high frequency band edge, and (b)
represents the average value of ${\rm c}_{0} / v_{\rm g}$ in the
gap. The circles in these figures correspond to the measurements,
while the open triangles to the FDTD simulations. The group velocity
at the band edge is rapidly reduced as the thickness of the crystal
increases. With only four unit cells this group velocity is almost a
factor 15 slower than the speed of light in vacuum. On the other
hand, the group velocity in the gap rapidly increases as the
thickness of the crystal is augmented. For a photonic crystal of 3
unit cells the group velocity is roughly two times larger than the
speed of light in vacuum. The transmission at gap frequencies of the
thickest photonic crystal was lower than the noise level of our
setup. For an infinite and defect-free photonic crystal one should
expect a zero group velocity at the band edges and a no transmission
in the gap.

\section{summary}

Through the direct access of amplitude and phase, terahertz
time-domain spectroscopy offers unprecedented possibilities for the
fundamental study of the propagation of electromagnetic waves in
photonic structures. We have investigated the propagation of
single-cycle THz pulses through strongly scattering photonic
crystals with different thickness over a large bandwidth covering
more than an octave below, at and above the bandgap. At frequencies
within the gap we observe an attenuation length smaller than one
unit cell and anomalous dispersion. The strong dispersion introduced
by the crystal gives rise to extremely low group velocities at the
band edges and to superluminal transmission in the gap.

\section*{acknowledgements}

We gratefully acknowledge the assistance of C. Schotsch and K.
Berdel with the fabrication and characterization of the photonic
crystals. We are indebted with A. Stahl for fruitful discussions and
with F. Merget and O. Winkler for the scanning electron microscope
photographs. We acknowledge financial support by the European
Commission under the project {\it Interaction}.


\begin{thebibliography}{100}

\bibitem{Steinber:93}
A.M. Steinberg, P.G. Kwiat, and R.Y. Chiao, {\it Phys. Rev. Lett.}
{\bf 71}, 708 (1993).

\bibitem{Spielmann:94}
Ch. Spielmann, R. Szip\"{o}cs, A. Stingl, and F. Krausz, {\it Phys.
Rev. Lett.} {\bf 73}, 2308 (1994).

\bibitem{Longhi:01}
S. Longhi, M. Marano, P. Laporta, and M. Belmonte, {\it Phys. Rev.
E} {\bf 64}, 055602(R) (2001).

\bibitem{Hache:02}
A. Hach\'{e} and L. Poirier, {\it Appl. Phys. Lett.} {\bf 80}, 518
(2002).

\bibitem{Munday:02}
J.N. Munday and W.M. Robertson, {\it Appl. Phys. Lett.} {\bf 81},
2127 (2002).

\bibitem{Solli:03}
D.R. Solli, C.F. McCormick, C. Ropers, J.J. Morehead, R.Y. Chiao,
and J.M. Hickmann, {\it Phys. Rev. Lett.} {\bf 91}, 143906 (2003).

\bibitem{Galli:04}
M. Galli, B. Bajoni, F. Marabelli, L.C. Andreadi, L. Pavesi, and G.
Pucker, {\it Phys. Rev. B} {\bf 69}, 115107 (2004).

\bibitem{Tarhan:95}
\.{I}.\.{I}. Tarhan, M.P. Zinkin, and G.H. Watson, {\it Opt. Lett.}
{\bf 20}, 1571 (1995).

\bibitem{Imhof:99}
A. Imhof, W.L. Vos, R. Sprik, and A. Lagendijk, {\it Phys. Rev.
Lett.} {\bf 83}, 2942 (1999).

\bibitem{Vlasov:99}
Y.A. Vlasov, S. Petit, G. Klein, B. H\"{o}nerlage, and Ch.
Hirlimann, {\it Phys. Rev. B} {\bf 60}, 1030 (1999).

\bibitem{Bayindir:00}
M. Bayindir, B. Temelkuran, and E. Ozbay, {\it Phys. Rev. B} {\bf
61}, R11855 (2000).

\bibitem{Notomi:01}
M. Notomi, K. Yamada, A. Shinya, J. Takahashi, C. Takahashi, and I.
Yokohama, {\it Phys. Rev. Lett.} {\bf 87}, 253902 (2001).

\bibitem{Wiersma:97}
D. S. Wiersma, P. Bartolini, A. Lagendijk, and R. Righini, {\it
Nature} {\bf 390}, 671 (1994).

\bibitem{Noda:03}
B.-S. Song, S. Noda, and T. Asano, {\it Science} {\bf 300}, 1537
(2003).

\bibitem{Painter:99}
O. Painter, R.K. Lee, A. Scherer, A. Yariv, J.D. O'Brien, P.D.
Dapkus, and I. Kim, {\it Science} {\bf 284}, 1819 (1999).

\bibitem{Alivisatos:04}
P. Alivisatos, {\it Nature biotechnology} {\bf 22}, 47 (2004).

\bibitem{Lodahl:04}
P. Lodahl, A.F. van Driel, I.S. Nikolaev, A. Irman, K. Overgaag, D.
Vanmaekelberg, and W.L. Vos, {\it Nature} {\bf 430}, 654 (2004).

\bibitem{Yablonovitch:87}
E. Yablonovitch, {\it Phys. Rev. Lett.} {\bf 58}, 2059 (1987).

\bibitem{John:87}
S. John, {\it Phys. Rev. Lett.} {\bf 58}, 2486 (1987).

\bibitem{Soljacic:04}
M. Solja\v{c}i\'{c} and J.D. Joannopoulos, {\it Nature Materials}
{\bf 3}, 211 (2004).

\bibitem{Chiao:97}
R.Y. Chiao and A.M. Steinberg in Progress in Optics XXXVII, pp.
345-405, edited by E. Wolf (Elsevier Science B.V., 1997).

\bibitem{Nimtz:03}
G. Nimtz, {\it Prog. Quant. Elec.} {\bf 27}, 417 (2003).

\bibitem{Stenner:03}
M.D. Stenner, D.J. Gauthier, and M.A. Niefeld, {\it Nature} {\bf
425}, 695 (2003).

\bibitem{Nimtz:04}
G. Nimtz, {\it Nature} doi:10.1038/nature02586 (2004).

\bibitem{Chelnokov:97}
A. Chelnokov, S. Rowson, J.-M. Lourtioz, L. Duvillaret, and J.-L.
Coutaz, {\it Electron. Lett.} {\bf 33}, (1997).

\bibitem{Gonzalo:02}
R. Gonzalo, B. Martinez, C.M. Mann, Harm Pellemans, P. Haring
Bolivar, and P. de Maagt {\it IEEE trans. Microwave Theory and
Techn.} {\bf 50}, 2384 (2002).

\bibitem{Ho:94}
K.M. Ho, C.T. Chan, C.M. Soukoulis, R. Biswas, and M. Sigalas, {\it
Solid St. Comm.} {\bf 89}, 413 (1994).

\bibitem{Sozuer:94}
H.S. Sozuer and J.P. Dowling, {\it J. Mod. Opt.} {\bf 41}, 231
(1994).

\bibitem{Exter:90}
M. van Exter and D. Grischkowsky, {\it Appl. Phys. Lett.} {\bf 56},
1694 (1990).

\bibitem{Ozbay:96}
E. \"{O}zbay, {\it J. Opt. Soc. Am. B.} {\bf 13}, 1945 (1996).

\bibitem{Lin:98}
S.Y. Lin, J. G. Fleming, D. L. Hetherington, B. K. Smith, R. Biswas,
K. M. Ho, M. M. Sigalas, W. Zubrzycki, S. R. Kurtz, and Jim Bur,
{\it Nature} {\bf 394}, 251 (1998).

\bibitem{Robertson:92}
W.M. Robertson, G. Arjavalingam, R.D. Meade, K.D. Brommer, A.M.
Rappe, and J.D. Joannopoulos, {\it Phys. Rev. Lett.} {\bf 68}, 2023
(1992).

\bibitem{Ozbay:94}
E. \"{O}zbay, E. Michel, G. Tuttle, R. Biswas, K.M. Ho, J. Bostak
and D.M. Bloom, {\it Opt. Lett.} {\bf 19}, 1155 (1994).

\bibitem{Jackson}
J.D. Jackson, {\it Classical electrodinamics}, chap. 7 (John Wiley
\& Sons, Inc., New York, 1999).

\bibitem{Capus:03}
C. Capus and K. Brown, {\it J. Acoust. Soc. Am.} {\bf 113}, 3253
(2003).

\bibitem{Gamble:99}
L.J. Gamble, W.M. Diffey, S.T. Cole, R.L. Fork, D.K. Jones, T.R.
Nelson, J.P. Loehr, and J.E. Ehret, {\it Opt. Express} {\bf 5}, 267
(1999).

\end{thebibliography}



\end{document}